\documentclass[11pt]{article}
\usepackage{amsfonts}
\usepackage{amsmath}
\usepackage[dvips]{epsfig}

\begin{document}

\author{A. de Souza Dutra$^{a,b}$\thanks{%
Corresponding author: E-mail: dutra@feg.unesp.br \ \ Phone:0055-12-31232848}%
, Marcelo Hott$^{a}$\thanks{%
E-mail: hott@feg.unesp.br} and Filipe F. Bellotti$^{a}$ \\
$^{a}$UNESP Univ Estadual Paulista, Campus de Guaratinguet\'{a}, \\
Departamento de F\'{\i}sica e Qu\'{\i}mica\thanks{%
Permanent Institution}\\
12516-410 Guaratinguet\'{a}, SP, Brasil.\\
$^{b}$Abdus Salam ICTP, Strada Costiera 11, 34014 Trieste Italy. }
\title{{\LARGE A mapping function approach applied to some classes of
nonlinear equations}}
\maketitle

\begin{abstract}
In this work, we study some models of scalar fields in 1+1 dimensions with
non-linear self-interactions. Here, we show how it is possible to extend the
solutions recently reported in the literature for some classes of nonlinear
equations like the nonlinear Klein-Gordon equation, the generalized
Camassa-Holm and the Benjamin-Bona-Mahony equations. It is shown that the
solutions obtained by Yomba \cite{yomba1}, when using the so-called
auxiliary equation method, can be reached by mapping them into some known
nonlinear equations. This is achieved through a suitable sequence of
translation and power-like transformations. Particularly, the parent-like
equations used here are the ones for the $\lambda \phi ^{4}$ model and the
Weierstrass equation. This last one, allow us to get oscillating solutions
for the models under analysis. We also systematize the approach in order to
show how to get a larger class of nonlinear equations which, as far as we
know, were not taken into account in the literature up to now.

\medskip


\end{abstract}

\pagebreak \newpage \topmargin -1cm

In the last few years, a growing number of works have been devoted to obtain
novel analytical solutions for some classes of nonlinear differential
equations, like the nonlinear Klein-Gordon equation, the generalized
Camassa-Holm and Benjamin-Bona-Mahony equations. For this, many approaches
were developed. Here we intend to show that through a sequence of
transformations, were one alternates translations and power-like
transformations, some of the solutions reported recently in the literature
can be reproduced and, better, increased. In particular, we treat the cases
studied by E. Yomba in a recent work \cite{yomba1}, as well as by J. Nickel
\cite{nickel}. In those works, the authors introduced interesting methods in
order to deal essentially with an auxiliary equation like

\begin{equation}
\left( \frac{dF(s)}{ds}\right) ^{2}=\sum_{i=0}^{N}h_{i}\,F^{i}.  \label{eq1}
\end{equation}

The idea is to show that those solutions can be obtained from a direct
mapping with an already known equation like, for instance, the one coming
from the so-called $\lambda \phi ^{4}$ model.

Let us begin with the Cases 1 and 2 of Yomba \cite{yomba1}. Starting with
the equation

\begin{equation}
\left( \frac{d\phi(s)}{ds} \right)^2=A\,\phi^4+B\,\phi^2+C,  \label{eq2}
\end{equation}

\noindent and performing the transformation $\phi=\left(F^{-2}+\beta%
\right)^{-\frac{1}{2}}$, we arrive at

\begin{equation}
\left( \frac{dF(s)}{ds} \right)^2=C+\left(B+3\beta
C\right)F^2+\left(A+2\beta B+3\beta^2
C\right)F^4+\beta\left(A+\beta\,B+\beta^2\,C\right)F^6,  \label{eq3}
\end{equation}

\noindent which is clearly written in the form appearing in the work \cite%
{yomba1}, namely

\begin{equation}
\left( \frac{dF(s)}{ds} \right)^2=h_0+h_2\, F^2+h_4\, F^4+h_6\, F^6.
\label{eq4}
\end{equation}

At this point we should observe that, in order to make contact with the
solution appearing in \cite{yomba1}, we must fix the arbitrary parameter $%
\beta $ such that $\beta =\frac{3\,h_{4}}{8\,h_{2}}$. Once this particular
choice is made, the restrictions appearing in the Case 1 of \cite{yomba1} ($%
h_{0}=\frac{8\,h_{2}^{2}}{27\,h_{4}},\,h_{6}=\frac{h_{4}^{2}}{4\,h_{2}}$)
are recovered. So, one can conclude that the freedom introduced by means of
the arbitrary parameter $\beta $, allows us to get less restrictive
solutions.

In fact, as we are going to see below, this mapping allows to get, besides
the trigonometric and hyperbolic solutions, some other oscillating ones
coming from the solutions of the Weierstrass equation \cite{whittaker}, \cite%
{abram}. Furthermore, as it can be seen, if one chooses the particular case
where $C=0$ (Case 2) in the equation (\ref{eq2}), one obtains the relations,

\begin{equation}
B= h_2, \, A= \epsilon \,\sqrt{ \Delta }, \, \beta= \frac{h_4+\epsilon\,
\sqrt{ \Delta }}{2\, h_2},  \label{eq5}
\end{equation}

\noindent where $\epsilon =\pm 1$ and $\Delta =h_{4}^{2}-4\,h_{2}h_{6}$, as
defined in the Table 2 of \cite{yomba1}. Note that, due to the restricted
choice of the parameter $C$ ($C=0$), the solutions of equation (\ref{eq2})
shall be given by

\begin{equation}
\phi(s)=\frac{2\,h_2\, e^{\pm\sqrt{h_2}s}}{1-\epsilon\,h_2\,\Delta\,e^{\pm 2%
\sqrt{h_2}s}},  \label{eq6}
\end{equation}

\noindent and the solution for the function $F(s)$ is simply obtained by
direct substitution in the equation $F_{(2)}=\pm \left( \phi ^{2}-\beta
\right) ^{-1/2}$.

Before going further in our analysis, we should mention that the Case 5 of
\cite{yomba1} can be obtained from this last case treated above, by making
the additional transformation $F_{(2)}=\pm F_{(5)}^{1/2}$. Then, we obtain
the following equation

\begin{equation}
\left( \frac{dF_{(5)}(s)}{ds}\right)
^{2}=h_{2(5)}\,F_{(5)}^{2}+h_{3(5)}\,F_{(5)}^{3}+h_{4(5)}\,F_{(5)}^{4},
\label{eq4b}
\end{equation}

\noindent where one shall make the identifications: $h_{2(5)}=4\,h_{2(2)}$, $%
h_{3(5)}=4\,h_{4(2)}$ and $h_{4(5)}=4\,h_{6(2)}$. Thus, we see that the
\emph{discriminant} defined for this case comes from the one of the Case 2
and is given by $\Delta _{(5)}=h_{3(5)}^{2}-4\,h_{2(5)}\,h_{4(5)}=16~\Delta
_{(2)}$. The notation $h_{i(j)}$ stands for the i-th coefficient of the j-th
case.

Now, we define \textquotedblleft generations\textquotedblright\ of equations
which can be obtained from the important nonlinear Weierstrass differential
equation

\begin{equation}
\left( \frac{d\wp }{ds}\right) ^{2}=4\,\wp ^{3}-g_{2}\,\wp -g_{3},
\label{eq7}
\end{equation}

\noindent which, as it is known, can present solutions in many regimes,
according to the specific choice of the \emph{invariants} $g_{2}$ and $g_{3}$
and of the \emph{discriminant} $\Delta ={g_{2}}^{3}-27{g_{3}}^{2}$. Those
solutions include, trigonometric and hyperbolic solutions as well as other
oscillating solutions in terms of the Jacobi elliptic functions.

What we call the \textquotedblleft zero-th\textquotedblright\ generation is
trivially obtained from the Weierstrass equation by simply performing a
translation and a dilation like $F_{0}(s)=a\,\wp (s)+b$. In this generation,
one has a quadratic term in the differential equation. One can check that
the Case 4 of \cite{yomba1} is completely recovered, including the same
constraints relating  the coefficients of the polynomial on the right hand
side of the equation.

The next generation is obtained by doing the transformation $\wp
(s)=r\,(F_{I})^{\alpha }$ ($r$ and $\alpha $ are arbitrary constants) and,
if one is interested only in polynomials with positive and integer degrees
in $F_{I}$, one must choose $\alpha =-1$ or $\alpha =-2$. In such cases, one
obtains:

\begin{equation}
\left( \frac{dF_{Ia}(s)}{ds} \right)^2=4\, r\,F_{Ia} -\frac{g_2}{r}%
\,F_{Ia}^3-\frac{g_3}{r^2}\,F_{Ia}^4, \,\,\, \mathrm{for} \, \alpha=-1 ,
\label{eq8}
\end{equation}

\noindent and

\begin{equation}
\left( \frac{dF_{Ib}(s)}{ds} \right)^2=r-\frac{g_2}{4\, r}F_{Ib}^4-\frac{g_3%
}{4\, r^2}F_{Ib}^6, \,\,\, \mathrm{for} \, \alpha=-2 .  \label{eq9}
\end{equation}

The second generation comes from a translation performed on the functions of
the first generation. For instance, by doing the transformation $%
F_{Ia}(s)=F_{IIa}(s)+k$ one is left with

\begin{equation}
\left( \frac{dF_{IIa}(s)}{ds} \right)^2=h_0+h_1\, F_{IIa}+h_2\,
F_{IIa}^2+h_3\, F_{IIa}^3+h_4\, F_{IIa}^4,  \label{eq10}
\end{equation}

\noindent where

\begin{eqnarray}
h_{0} &=&-\frac{k}{r^{2}}\left( g_{3}k^{3}+g_{2}\,r\,k^{2}-4\,r^{3}\right)
,~h_{1}=\frac{1}{k\,r^{2}}\left(
4\,r^{3}-4\,k^{3}g_{3}-3\,r\,k^{2}g_{2}\right) ,  \notag \\
h_{2} &=&-\frac{1}{r^{2}}\left[ 3\,k\left( 2\,k\,g_{3}+r\,g_{2}\right) %
\right] ,~h_{3}=-\frac{1}{r^{2}}\left( 4\,k\,g_{3}+r\,g_{2}\right) ~\mathrm{%
and~}h_{4}=-\frac{g_{3}}{r^{2}}.  \notag  \label{eq11}
\end{eqnarray}

\noindent It is evident that we have no longer the coefficients independent
of each other. At this point we can make the identifications defined in \cite%
{yomba1}, $\alpha =h_{4},\,\beta =\frac{h_{3}}{4},\,\gamma =\frac{h_{2}}{6}%
,\,\epsilon =h_{0}$, and verify that the constraints appearing in the Case 3
of \cite{yomba1} and also in \cite{nickel}, namely

\begin{equation}
g_{2}=\alpha \,\epsilon -4\,\beta \,\delta +3\,\gamma ^{2}~\mathrm{and~}%
g_{3}=\alpha \,\gamma \,\epsilon +2\,\beta \,\gamma \,\delta -\alpha
\,\delta ^{2}-\gamma ^{3}-\epsilon \,\beta ^{2}  \notag  \label{eq12}
\end{equation}

\noindent are reproduced and, as a consequence, the discriminant of the
system will be necessarily the same one for the Weierstrass equation: $%
\Delta =g_{2}^{3}-27\,g_{3}^{2}$. In this case the solution of the functions
$F_{IIa}$ are written in terms of the Weierstrass function as $%
F_{IIa}(s)=r\wp ^{-1}(s)-k$.

Finally, the Case 6 of \cite{yomba1} is precisely the equation (\ref{eq2}),
which we have commenced the mapping of the first two cases of \cite{yomba1}.

Now, we show that, for the cases where the three constants $A,B,C$ are
arbitrary, equation (\ref{eq2}) can be mapped into the Weierstrass equation.
For this we can set $\phi (s)=\sqrt{C}\,\left( \wp (s)-\frac{B}{3}\right) ^{-%
\frac{1}{2}}$ or $\phi (s)=\frac{1}{\sqrt{A}}\left( \wp (s)-\frac{B}{3}%
\right) ^{\frac{1}{2}}$, leading us to the equation (\ref{eq7}) with

\begin{equation}
g_{2}=4\, C \left(\frac{B^{2}}{3\, C}-A\right) \,\mathrm{and}\,~g_{3}= 4\,
C\, B\left( \frac{A}{3}-\frac{2\,B^{2}}{27\, C}\right) .  \label{eq13}
\end{equation}

\noindent for the first case or

\begin{equation}
g_{2}=4\, A \left(\frac{B^{2}}{3\, A}-C\right) \,\mathrm{and}\,~g_{3}= 4\,
A\, B\left( \frac{C}{3}-\frac{2\,B^{2}}{27\, A}\right) .
\end{equation}

\noindent in the second case. In fact, if one is interested in non-singular
solutions, one must choose the first solution, since the Weierstrass
function always presents singular points which, in that case, will lead to
well-behaved solutions. In the Figure 1, some examples of those continuous
solutions are presented.

Since we have shown how to obtain all the cases studied in \cite{yomba1} by
means of a mapping approach, we can go further with the procedure by getting
novel equations whose solution can be reached by using this method. In fact,
we can obtain terms of higher degrees polynomial on the right hand side of
the auxiliary differential through a suitable combination of power-like and
translation transformations on the Weierstrass differential equation.
However, in order to achieve this goal, one must choose the translation
constant appropriately. For example, let us start with the Weierstrass
equation and make the transformation $\wp (s)=(k+F(s)^{2})^{-2}$, which
leads us to the equation

\begin{equation}
\left( \frac{dF(s)}{ds}\right)
^{2}=a_{0}+a_{2}\,F^{2}+a_{4}\,F^{4}+a_{6}\,F^{6}+a_{8}\,F^{8}+a_{10}%
\,F^{10},  \label{eq14}
\end{equation}

\noindent where we have fixed the invariant $g_{3}$ as $g_{3}=\frac{4\ -\
g_{2}\,k^{4}}{k^{6}}$ in order to avoid a term with negative power on the
right hand side of the differential equation. Under this condition, the
coefficients $a_{i}$ are given by

\begin{eqnarray}
a_{0} &=&(-12+g_{2}k^{4})/(8\,
k),\,~a_{2}=3(-20+3g_{2}k^{4})/16k^{2},~\,a_{4}=(-5+g_{2}k^{4})/k^{3},\,
\notag \\
a_{6}
&=&(-30+7g_{2}k^{4})/8k^{4},\,~a_{8}=3(-4+g_{2}k^{4})/8k^{5},%
\,~a_{10}=(-4+g_{2}k^{4})/16k^{6}.  \label{eq15}
\end{eqnarray}

\noindent Note that, even if we include an additional parameter through a
scaling of the function, we still have only three free parameters, the
remaining ones should be fixed in order to grant the exact solvability of
the equation. Some solutions for this example are illustrated in the Figure
2.

One can see that the approach developed here generates higher order exactly
solvable differential equations. We think that those can be useful to a
better understanding of some classes of nonlinear equations like the
generalized Camassa-Holm and the Benjamin-Bona-Mahony equations, as done by
Yomba \cite{yomba1} and the nonlinear Klein-Gordon equation and their
generalizations as the sine-Gordon, the sinh-Gordon and their modified
versions, as considered by \cite{wazwaz}.

In order to illustrate the mapping approach even more, we would like to
consider here non-linear models involving the Jacobi elliptic functions, \
particularly what has been called the Jacobi model \cite{dunne}, whose
auxiliary differential equation is%
\begin{equation}
\left( \frac{d\phi }{ds}\right) ^{2}=4~sn^{2}(\frac{\phi }{2}|m)-4c^{2},
\label{eq16}
\end{equation}%
where $c$ is a constant of the integration and $sn(\varphi |m)$ is the
Jacobi elliptic function defined by $sn(\varphi ,m)=\sin \theta $, with
respect to the integral $\varphi =\int\limits_{0}^{\theta }d\alpha /(1-m\sin
^{2}\alpha )^{1/2}$ , where $\theta $ is called the \textit{amplitude} and $%
m $ ($0\leq m\leq 1$) is the \textit{elliptic parameter}. The equation (\ref%
{eq16}) retrieves the one for the so-called sine-Gordon model for $m=0$. We
have noted, by using the mapping approach, that the above equation has
oscillating solutions for $0<c^{2}<1\,$\ and soliton solutions for $c^{2}=0$%
. The redefinition of the fields $F(s)=cn(\varphi (s)|m)$, where $cn(\varphi
|m)$ is the Jacobi elliptic function defined by $cn(\varphi |m)=\cos \theta $%
, maps the equation (\ref{eq16}) into an equation identical to (\ref{eq1})
with $N=6$ and the coefficients given by
\begin{equation}
h_{0}=1-m-c^{2}+c^{2}m,~h_{2}=-2+c^{2}+3m-2c^{2}m,~h_{4}=1-3m+c^{2}m,~h_{6}=m.
\label{eq17}
\end{equation}

By using the following mapping%
\begin{equation}
F=\pm \sqrt{h_{0}}(\wp +\kappa )^{-1/2}  \label{eq18}
\end{equation}%
with $\kappa =1/3(2-c^{2}-3m+2c^{2}m)$, we recover the Weierstrass
differential equation (\ref{eq7}) satisfied by the Weierstrass elliptic
function $\wp (s;g_{2},g_{3})$ with the invariants%
\begin{eqnarray}
g_{2} &=&4/3(1-c^{2}(1+m)+c^{4}(1-m+m^{2})),  \notag \\
g_{3} &=&-4/27(2-3c^{2}(1+m)-3c^{4}(1-4m+m^{2})+c^{6}(2-3m-3m^{2}+2m^{3}).
\label{eq19}
\end{eqnarray}

At this point we can use the relation 18.9.11 of the reference \cite{abram},
which relates the Weierstrass and the Jacobi elliptic functions, since the
discriminant $\Delta >0$. Then we obtain%
\begin{equation}
\wp (s)=e_{3}+(e_{1}-e_{3})/sn^{2}((e_{1}-e_{3})^{1/2}s|\nu ),  \label{eq20}
\end{equation}%
where%
\begin{equation}
e_{1}=1/3(1+c^{2}-2c^{2}m),~e_{2}=1/3(1-2c^{2}+c^{2}m),~e_{3}=1/3(-2+c^{2}+c^{2}m)
\label{eq21}
\end{equation}%
and%
\begin{equation}
\nu =(e_{2}-e_{3})/(e_{1}-e_{3})=(1-c^{2})(1-c^{2}m).  \label{eq22}
\end{equation}

Finally the solution for the Jacobi elliptic model can be written as%
\begin{equation}
\phi _{Jacobi}(s)=2\ cn^{-1}(F(s)|m),  \label{eq23a}
\end{equation}%
with%
\begin{equation}
F(s)=\pm \frac{\sqrt{(1-c^{2})(1-m)}sn((1-c^{2}m)s|\nu )}{\sqrt{%
(1-c^{2}m)-(1-c^{2})m~sn^{2}((1-c^{2}m)s|\nu )}}.  \label{eq23b}
\end{equation}

The above expression is the solution for the equation (\ref{eq1})~with $N=6$
and coefficients $h_{i}$ given in (\ref{eq17}). According to the
classification scheme mentioned above, it belongs to the 2nd generation ($%
F_{IIb}$). By taking $c=0$ in equations (\ref{eq22})-(\ref{eq23b}) we obtain
the \textit{(anti-)kink soliton} solutions for the Jacobi elliptic model%
\begin{equation}
\phi _{Jacobi}(s)=2\ cn^{-1}\left( \pm \frac{\sqrt{(1-m)}\tanh s}{\sqrt{%
1-m\tanh ^{2}s}}\left\vert
\begin{array}{c}
m \\
\\
\end{array}%
\right. \right)  \label{eq23c}
\end{equation}%
The \textit{kink} solution above can be rewritten as $\phi
_{kink-Jacobi}(s)=2K[m]+2~sn^{-1}(\tanh s|m)$, which is the \textit{kink}
found in \cite{dunne}, and $K[m]$ is the elliptic quarter period.

The solutions for the sine-Gordon model are retrieved by taking $m=0$. In
this limit the differential equation for $F\left( s\right) $ is the equation
(\ref{eq1}) with $N=4$; the coefficients can be obtained from (\ref{eq17})
by taking $m=0$. The solution for the sine-Gordon model is%
\begin{equation}
\phi _{s-G}(s)=2\cos ^{-1}(\pm \sqrt{1-c^{2}}sn(s|1-c^{2})),  \label{eq24}
\end{equation}%
which is in agreement with the solution presented in \cite{japa} and \cite%
{riva} with an adequate reparametrization of the variable $s$ and a
reescaling of the field. For $c=0$ we obtain the \textit{(anti-)kink}
solution for the sine-Gordon model%
\begin{equation}
\phi _{s-G}(s)=2\cos ^{-1}(\pm \tanh s)=4\tan ^{-1}(e^{\mp s}).  \label{eq25}
\end{equation}

Finally we would like to mention that generalizations of non-polynomial
non-linear models as for instance, the generalized versions of the
sine-Gordon and the sinh-Gordon models \cite{wazwaz}, can also be approached
by means of this mapping function method described here.

\bigskip

\noindent\textbf{Acknowledgments:} The authors thanks to CNPq and FAPESP for
partial financial support. This work was partially done during a visit
(A.S.D.) within the Associate Scheme of the Abdus Salam ICTP.

\newpage

\begin{figure}[tbp]
\begin{center}
\begin{minipage}{20\linewidth}
\epsfig{file=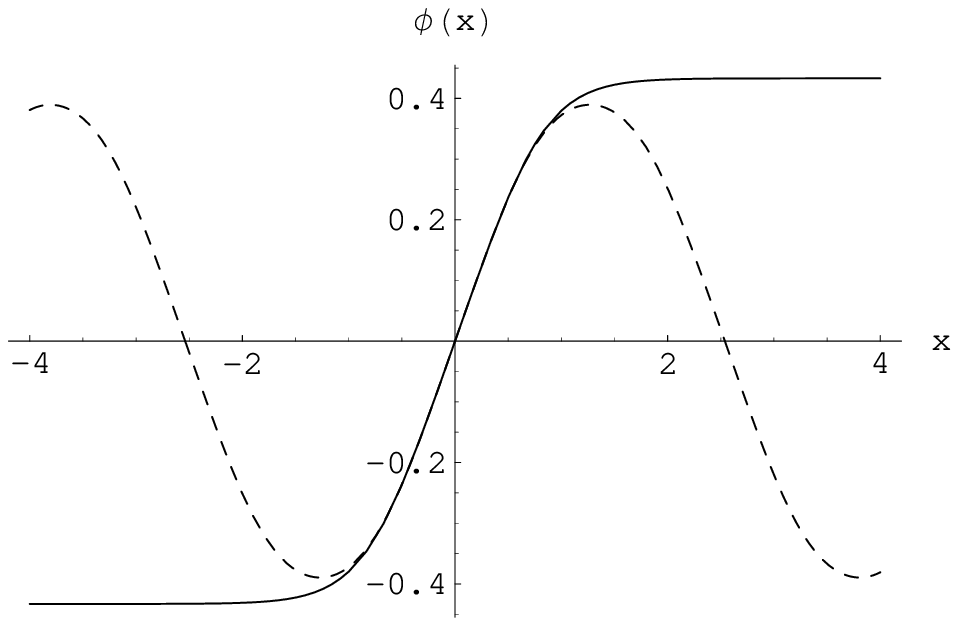}
\end{minipage}
\end{center}
\caption{Solutions of the equation (2) with $\Delta=0$, $g_2=12$ and $g_3=-8$
(solid line) and $\Delta > 0$, $g_2=13$ and $g_3=-8$ (dashed line)}
\label{fig:fig1}
\end{figure}

\begin{figure}[tbp]
\begin{center}
\begin{minipage}{20\linewidth}
\epsfig{file=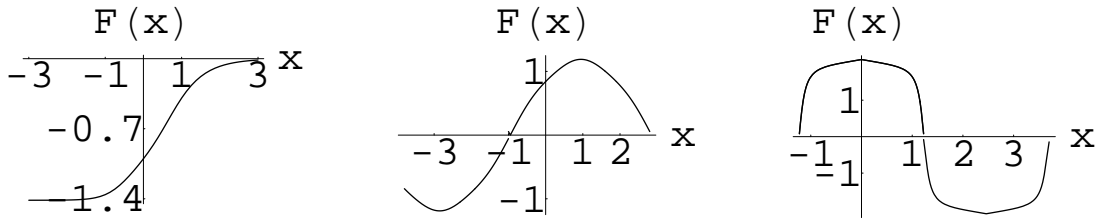,width=160mm}
\end{minipage}
\end{center}
\caption{Solutions of the equation (14) for $\Delta=0$, $g_2=12$ and $g_3=-8$
(left figure); $\Delta > 0$, $g_2=20$ and $g_3=-8$ (middle figure) and $%
\Delta < 0$, $g_2=-20$ and $g_3=1$ (right figure)}
\label{fig:fig2}
\end{figure}

\end{document}